\documentclass[conference]{IEEEtran}
\IEEEoverridecommandlockouts

\usepackage{cite}
\usepackage{amsmath,amssymb}
\usepackage{graphicx}
\usepackage{xcolor}
\usepackage[hidelinks]{hyperref}
\usepackage{xurl}

\newcommand{\seedhint}[1]{\textit{\textcolor{gray}{#1}}}

\begin{document}

\title{Agent Trajectories as Evidence for Software Quality Assurance}

\author{
    \IEEEauthorblockN{Jonathan Ullrich, Julien Siebert, Michael Kläs, Adam Trendowicz, Andreas Jedlitschka}
    \IEEEauthorblockA{Fraunhofer IESE \\
    Kaiserslautern, Germany \\
    \{first.last\}@iese.fraunhofer.de}
}

\maketitle

\begin{abstract}
Autonomous software agents increasingly plan, implement, and test software with limited human oversight. Their decisions may introduce implicit assumptions that final code and passing tests do not adequately expose. We propose analyzing observable agent trajectories alongside software and agent specifications and AI generated artifacts as evidence for targeted quality assurance. The aim is to prioritize consequential decisions for scrutiny, investigate their possible contribution to software problems, and inform improvements to specifications and development practices. Industrial experience motivates the idea; available execution records and preliminary analysis methods make it plausible. However, trajectories alone do not establish that decisions are justified, outcomes acceptable, or corrective changes effective. We ask how this evidence can support review, diagnosis, and continuous quality improvement, and when simpler QA approaches suffice. Perspectives from requirements engineering, verification and validation, agent systems, causal inference, empirical software engineering, and human factors could help develop and evaluate the idea. 
\end{abstract}


\section{The idea}

Autonomous software agents can make consequential development decisions without explicit human review. They may resolve ambiguous requirements, introduce assumptions, and implement behavior that was never explicitly specified. Tests generated within the same workflow may reproduce these assumptions rather than challenge them. When the software behaves unexpectedly, generated documentation, code and tests alone may not explain whether the problem arose from an ambiguous requirement, an unsupported assumption, an implementation error, or the conditions under which the agent operated.

We propose treating observable \emph{agent trajectories}~\cite{10.1145/3798271} as first-class software engineering artifacts for quality assurance. Trajectories include recorded reasoning outputs, tool calls, context access, artifact changes, and verification activity. Analyzed alongside software specifications, agent specifications, and AI generated artifacts, they may help reviewers identify consequential decisions, assess their justification, and investigate their contribution to software problems. Here, agent specifications include instructions, reusable skills, tool-use constraints, and harness configuration. Findings could inform changes to these specifications and to development and verification practices, forming a continuous feedback loop.

\section{Why it matters now}

Agentic development is moving toward workflows in which agents plan, implement, test, and revise software with limited human involvement. In such ``dark factory'' settings,\footnote{\url{https://www.openhands.dev/blog/what-is-an-ai-dark-factory}} developers may have less direct knowledge of the decisions behind an implementation (a phenomenon described as \emph{cognitive debt} \cite{storey2026technicaldebtcognitiveintent}). Reviewing every execution step is impractical, while reviewing only final artifacts can leave important assumptions and development conditions unexplored. Quality assurance therefore needs ways to focus scrutiny and learn from problems without requiring humans to reconstruct every run manually.

More execution data does not automatically provide better assurance. A plausible explanation may lack external justification, and passing tests may share the assumptions of the implementation. Different problems can also require different responses: an implementation error, missing domain knowledge, ambiguous requirements, and inadequate verification should not all lead to the same corrective action.  Regulated development makes this challenge particularly visible because decisions and their justification may need to withstand independent review.

In several industrial projects, we have observed agents introduce simplifications, rely on incorrect assumptions, or proceed without relevant context. They may use incomplete or outdated information, leave placeholders or incomplete functionality, or unnecessarily reimplement existing functionality. Agents may also depart from established architectural decisions or introduce new ones without explicit review. Such decisions are not necessarily wrong, but their justification and consequences can be difficult to assess. Agent-generated tests may fail to reveal problems when they rely on the same assumptions as the implementation.

One of these projects concerns a migration from COBOL to Java involving more than one million lines of code. The legacy codebase was represented in a code knowledge graph, and agents were equipped with skills to query that graph during reimplementation. The graph also supported parity checks between the legacy and migrated implementations. Such settings, when available, provides concrete sources against which agent decisions can be examined: legacy behavior, graph queries and their results, changes to the implementation, and the checks performed.

Trajectories could connect these sources to support both review of migration decisions and investigation of behavioral mismatches. For example, if migrated behavior differs from the legacy system, recorded queries, retrieved context, and artifact changes could help investigate whether relevant behavior was overlooked, incorrectly interpreted, or incorrectly implemented. These are competing explanations, not conclusions established by the records alone. Distinguishing them could guide different corrective actions: clarifying the migration specification, changing the agent's context-retrieval instructions, or strengthening parity checks.


\section{Why it is plausible}
Agent harnesses, such as OpenCode\footnote{\url{https://opencode.ai/}}, OpenHands\footnote{\url{https://www.openhands.dev/}}, or Claude Code\footnote{\url{https://code.claude.com/docs/en/overview}}, together with observability platforms, such as Langfuse\footnote{\url{https://langfuse.com/}} or Pydantic Logfire\footnote{\url{https://pydantic.dev/logfire}}, can already capture tool interactions, context retrieval, execution outputs, artifact changes, and test results. Version control and verification systems provide complementary records. Although monitoring varies across harnesses and configurations, these sources make it feasible to study agentic development as a recorded process rather than examining only its final products.

Preliminary work on analyzing agent execution records provides starting points for more targeted investigations~\cite{liu206plan}. Artifact histories may indicate where a problematic behavior first appeared, while preceding interactions may reveal the information available to the agent. Comparing this evidence with specifications, domain sources, and verification results could help reconstruct relevant decisions and formulate competing explanations. Existing automated review and verification approaches~\cite{otoum2026review, apostolou2026assistanceautonomysystematicliterature} provide starting points for scrutiny. Their effectiveness requires independent evaluation, particularly when review agents may share assumptions with the implementation agents.

Preliminary work on causal analysis \cite{siebert2023applications,giamattei2025causal} and counterfactual replay~\cite{shah2026causalagentreplaycounterfactual} also offers approaches for investigating agent executions and assessing interventions. The development workflow provides potential intervention points: software specifications, agent instructions, context-access mechanisms, and verification rules can be changed and evaluated in subsequent executions. Together, these research directions and available intervention points make investigation and feedback technically plausible. They do not yet establish under which conditions consequential decisions can be reliably recovered, causal explanations supported, or corrective changes shown to improve software quality.

\section{What we do not yet understand}
\textbf{Which decisions warrant scrutiny?}
Reviewing every agent decision is infeasible, yet prioritization must often happen before its consequences are fully understood. Candidate criteria include \emph{impact}, the potential consequences for correctness, safety, security, or compliance;  \emph{scope}, the extent to which a decision affects components and stakeholders; \emph{uncertainty}, gaps or conflicts in its supporting evidence; \emph{reversibility}, the difficulty and cost of undoing it and mitigating its consequences;  and \emph{validation effort}, the resources and expertise needed for independent assessment.

It remains unclear how these criteria should be assessed and combined. Even when validation is costly, a high-impact decision with weak supporting evidence may warrant scrutiny. The abstraction level alone cannot determine priority, since a local implementation decision may have system-wide consequences. Similarly, an agent's expressed confidence cannot substitute for supporting evidence. Beyond prioritization, we need to understand which decisions can be adequately assessed through independent automated checks and which require the judgment of stakeholders or domain experts, as well as what evidence supports that distinction.

\textbf{What evidence enables effective scrutiny?}
Relevant decisions may be distributed across various artifacts, including plans, retries, tool interactions, and artifact revisions. Each may reflect a different version of a decision. Accessing an authoritative source does not guarantee a correct interpretation. Likewise, the absence of a source link does not necessarily mean a decision is incorrect. Generated explanations may be incomplete or inaccurate accounts of how an outcome was produced. Therefore, we need to understand what evidence is sufficient for a particular review task and what information trajectory analysis provides beyond specifications, artifacts, and independent verification.

\textbf{When can analysis explain software problems?}
A reconstructed sequence of events may suggest an explanation, but it does not establish causation. Problems may arise from interacting decisions, confounded effects, incomplete specifications, tool behavior, and verification gaps, rather than from a single root cause. It remains unclear how to distinguish competing explanations with observational evidence, when interventions are necessary, and how nondeterminism or changing environments limit counterfactual analysis and replay. Therefore, the suitability of different methods must be investigated rather than assumed.

\textbf{When does feedback improve quality?}
A convincing explanation does not necessarily determine which corrective action is needed. Revising a software specification, changing agent instructions, or adding verification addresses different mechanisms and may introduce costs or regressions. We do not yet know how to select and validate these changes, whether the improvements can be generalized beyond the investigated case, or which parts of the feedback loop can be safely automated. As specifications, agents, and software evolve, evidence and conclusions must also remain reviewable.

\section{The question we would put to the room}

How might we analyze agent trajectories alongside software and agent specifications to prioritize consequential decisions, investigate software problems, and guide continuous quality improvement in autonomous development?

We would explore three connected questions:
\begin{itemize}
    \item Which decisions, including those based on implicit assumptions, warrant scrutiny, and which require human judgment?
    \item What evidence and analysis methods enable effective review and credible explanations of software problems?
    \item How can findings inform corrective changes, and how can we determine whether those changes improve quality?
\end{itemize}

\section{What would advance it}
Requirements engineering, quality assurance, quality modeling, and human factors could help establish criteria for prioritizing candidates and identify the evidence needed for different review tasks. An initial decision-and-evidence model could distinguish between assumptions introduced during the design process, implementation choices, observed problems, and proposed explanations. Structured representations could link decisions to specifications, sources, artifact modifications, and verification results~\cite{ullrich2025requirements}. This would support both proactive scrutiny and subsequent investigation.

Perspectives from multi-agent systems and causal inference can help us explore how observational analyses and interventions distinguish between competing explanations. Provenance analysis, artifact comparison, counterfactual analysis, controlled experiments, and replay are candidate approaches for addressing different investigative questions. We invite discussion on the feasibility and informativeness of these approaches, the assumptions they require, and the circumstances under which incomplete records, nondeterminism, or limited experimental control render the available evidence insufficient.

Workshop participants could also examine the circumstances under which these investigations provide useful assurance evidence. Determining whether a decision influenced an outcome is different from evaluating whether the outcome is acceptable. Therefore, independent tests, domain expertise, and stakeholder judgment may remain necessary. Similarly, an explanation alone does not establish an effective corrective action. Findings may inform changes to software specifications, agent instructions, configurations, or verification practices; however, those changes require validation. The benefits of investigations and feedback must be weighed against their costs and against simpler QA alternatives.

A human-factors perspective could help determine how to present an agent's progress and decisions in a timely manner for meaningful review. This includes determining whether the agent's approach to structuring the work is understandable to the team. Additionally, research is needed on how to maintain evidence and conclusions as code, specifications, and documentation evolve. These concerns link analytical methods to their practical application. Evidence must remain interpretable, relevant, and accessible while reviewers can still impact the development process.

Using existing research, experience reports, open-source infrastructure, and benchmarks, empirical software engineering could help shape candidate evaluation designs. Participants could consider what these settings reveal and which relevant tasks or development conditions they omit. Comparisons with QA based on specifications, code, tests, and conventional logs could determine the value of these additional trajectories. Outcomes could include identifying decisions, assessing diagnostic accuracy, evaluating review effort, and determining whether corrective changes prevent recurrence or improve quality in subsequent tasks without causing unacceptable regressions. Concrete workshop outcomes could include annotated examples, competing explanations, proposed interventions, and an initial evaluation design.


\bibliographystyle{IEEEtran}
\bibliography{references}

\end{document}